\documentclass[jou,a4paper,draftfirst, hidelinks]{apa7}

\usepackage{lipsum}

\usepackage[american]{babel}
\usepackage{csquotes}
\usepackage[style=apa,sortcites=true,sorting=nyt,backend=biber]{biblatex}
\usepackage{booktabs}
\usepackage{graphicx}
\usepackage{siunitx}

\DeclareLanguageMapping{american}{american-apa}
\title{Is a Mean Difference of 0.46 Relevant? Towards Determining the Smallest Effect Size of Interest for Visual Aesthetics of Websites}
\shorttitle{Smallest Effect Size of Interest for Visual Aesthetics of Websites}

\author{Florian Brühlmann}
\affiliation{University of Basel}

\leftheader{Brühlmann}

\abstract{Imagine conducting a study to determine whether users rate the visual aesthetics of your website more positively than your competitors. To assess users' perceptions of both websites, you use a validated survey scale for visual aesthetics, and you observe a statistically significant difference in users' ratings of the visual aesthetics of the two websites of 0.5 on a 7-point Likert-type scale. However, determining whether such a difference is practically (and theoretically) meaningful is challenging.
In this paper, I follow the procedure outlined in \textcite{anvariUsingAnchorbasedMethods2021}, to determine the smallest subjectively experienced difference in VisAWI-s ratings using an anchor-based method.
A sample of $N = 249$ participants rated and compared screenshots of eight websites in an online survey. 
I determined an estimate of a population-specific mean difference of 0.4, or in POMP units 6.58\%, which translates to a mean difference of 0.46 with the 7-point Likert-type scale of the VisAWI-s. These values suggest that differences in VisAWI-s scores exceeding these estimates, such as the 0.5 mentioned above, are likely noticeable and meaningful to users. However, the estimate of this smallest subjectively experienced difference is affected by the overall visual aesthetics rating of the stimuli used. 
Researchers can use this effect size to inform study design and sample size planning. Still, whenever possible, they should aim to determine a domain- and research-design-specific smallest effect size of interest.
}

\keywords{Smallest effect size of interest, Aesthetics, VisAWI, User experience, Practical significance, Minimum important difference, Smallest subjectively experienced difference}

\authornote{
   \addORCIDlink{Florian Brühlmann}{0000-0001-8945-3273}

\textbf{Important: This paper is a work-in-progress on NOT peer-reviewed. It is a collection of my ideas and analyses, which may become a paper at a later date. }

  Correspondence concerning this manuscript should be addressed to Florian Brühlmann, Faculty of Psychology, Center for General Psychology and Methodology, University of Basel, Missionsstrasse 62a, 4055 Basel.  E-mail: {florian.bruehlmann@gmail.com}}

\begin{document}
\maketitle
Visual aesthetics is essential for website owners, as users make quick decisions on whether to stay on a site or leave it \parencite[e.g., ][]{lindgaardAttentionWebDesigners2006}. Therefore, it is critical to measure users' perception of the visual aesthetics of a website. This is what the Visual Aesthetics of Websites Inventory (VisAWI) is designed for: The VisAWI is a survey scale measuring four facets of visual aesthetics (simplicity, diversity, colorfulness, and craftsmanship) with 18-items \parencite{moshagenFacetsVisualAesthetics2010}. In 2013, \textcite{moshagenShortVersionVisual2013} developed the VisAWI-s, a four-item short version of the scale. In practice, UX researchers may often need to rely on such short scales to measure different aspects of the user experience (e.g., usability with the UX-lite), to balance informativeness and participant burden in online or lab settings. 

However, what does a difference of 0.5 on the average of the four VisAWI-s items stand for? A statistically significant effect does necessarily mean that this effect is also practically and theoretically significant. This is especially apparent when considering big data research, which can provide such high power to statistical tests, that even minuscule differences can become statistically significant. Does that mean that participants will notice a difference? Will they spend more money, rate my website more favorably, or be more likely to recommend my website to friends? If a smallest effect size of interest can be determined, these questions can actually be answered even in cases where statistical tests have very high power. 
Another benefit of determining the smallest effect size of interest is that it makes it possible to design an informative and falsifiable study. If we can determine the smallest effect size of interest, we can choose a sample size that provides a statistical test with high power to detect it. We can also test for and demonstrate the absence of an effect that is large enough to be considered meaningful using equivalence tests \parencite{lakensEquivalenceTestingPsychological2018a}.

While researchers commonly rely on Cohen's (1988) benchmarks for small, medium, and large effect sizes, these are “arbitrary conventions, recommended for use only when no better basis for estimating the effect size is available” (\citeauthor[][]{cohenStatisticalPowerAnalysis1988}, 1988, p.12, in \textcite{anvariUsingAnchorbasedMethods2021}). It is often recommended to review existing literature to determine common effect sizes published in specific fields. While this is better than relying on arbitrary conventions, such reviews still do not establish which effect sizes are meaningful for specific research lines. Thus, determining the smallest effect size of interest that is specific to the research question and outcome measure is more appropriate.

One approach is to perform cost-benefit analyses and determine the size of an effect that could be considered beneficial enough to be worth the costs of a change. Using revenue data, it might be possible to estimate how VisAWI-s scores relate to revenue, and it could help to estimate how much revenue change a difference of 0.5 is related to. An alternative approach, conceptually related to a just noticeable difference in perception research, is the estimation of a minimum threshold of importance for self-reported outcomes. Here the goal is to determine a minimal detectable or minimal important difference. 
In User Experience (UX) research, this human-centered approach towards meaningful effect size interpretations could be applicable in various measures. To determine such a minimally important difference from a user perspective, the anchor-based method introduced in \textcite[]{anvariUsingAnchorbasedMethods2021} seems promising. 
%In UX research, a promising approach is the anchor-based method to determine minimally important differences between websites or website versions introduced in \parencite[]{anvariUsingAnchorbasedMethods2021}. 

Following their guidelines, I conducted a study to determine the meaning of differences in VisAWI-s scores for users. I let a sample of 249 participants rate 4 websites from the category News \& Media Publisher and 4 websites from the category Shopping \& E-commerce. 
Results show that an effect size of Cohen's $d_{z} = 0.26$, in percentage of maximum score units (POMP units) 6.58\%, or a difference on the 7-point Likert scale of the VisAWI-s of 0.46 is the smallest subjectively experienced difference using the stimuli in this study. 

Thus, differences in VisAWI-s ratings above these thresholds should correspond to a different aesthetic "experience" of users of the website. However, the estimate of the threshold may be biased towards the specific context examined (news and shopping websites from Australia). Since I asked users to rate different websites and not different versions of the same websites, it may overestimate the smallest subjectively experienced difference in VisAWI-s ratings for comparisons between different versions of the same websites. In the following, I outline the method I used to determine this effect size, how researchers may use this method in various contexts, what the limitations of this method are, and what implication results have for survey research.

\section{Method}
\subsection{Participants}
Based on the observed difference for the PANAS in \textcite{anvariUsingAnchorbasedMethods2021}, I expected an effect size of similar magnitude (d = .35) for the present study. Using the same calculations as \textcite{anvariUsingAnchorbasedMethods2021}, a sample size of about 500 participants in the "little difference" group (those participants who selected "little less" and "little more" in the comparison item) was estimated. However, since no previous data was available to estimate the number of participants falling in this group, I planned with a sample size of 250. This was within the budget for exploratory projects and since participants were required to compare 4 websites for two categories, I received 6 comparisons from each participant for each category of websites. I guessed that about 40\% of participants per comparison (approximately 100 participants) would fall into the "little difference" group, resulting in a sample of 600 observations (6 comparisons X 100 participants) for each website category. Participants were recruited from the United States using Prolific (\url{http://www.prolific.co}), with a required approval rate of at least 98, no more than 200 submissions, English-speaking skills and a balanced sample in terms of gender, using Prolific's built-in function. After excluding incomplete responses and two participants who commented that their data should not be used, a total of $N = 249$ participants (50\% female; age M = 37.0 years, SD = 13.4, range = 18 to 80 years) remained.

\subsection{Stimuli}
I selected websites from the categories News \& Media Publishers and Shopping \& E-commerce for the study. This was based on the assumption that the web presence of websites from these categories is likely relatively important for their revenue, making it more likely that they invest in a good user experience, which includes high visual aesthetics. To reduce potential brand or familiarity effects, websites from Australia that ranked high on the website similarweb\footnote{\url{https://www.similarweb.com/}}, but did not belong to the top 10, were selected. The URLs of the selected websites are presented in~\autoref{tab:websites}.

% \usepackage{booktabs}
% \usepackage{graphicx}
\begin{table}[]
\centering
\caption{Eight websites from Australia were selected as stimuli for the study.  }
\label{tab:websites}
\resizebox{\columnwidth}{!}{%
\begin{tabular}{@{}lllr@{}}
\toprule
Category                 & Website               & Url                                                                & Rank \\ \midrule
News \& Media Publishers & ABC                   & \url{https://www.abc.net.au/}            & 18   \\
 & 9News                 & \url{https://www.9news.com.au/}         & 45   \\
 & Sidney Morning Herald (SMH) & \url{https://www.smh.com.au/}             & 60   \\
 & Heraldsun             & \url{https://www.heraldsun.com.au/} & 143  \\
Shopping \& E-commerce   & Myer                  & \url{https://www.myer.com.au/}         & 33   \\
   & BigW                  & \url{https://www.bigw.com.au/}          & 39   \\
   & Catch                 & \url{https://www.catch.com.au/}     & 80   \\
   & Kogan                 & \url{https://www.kogan.com/au/}        & 109  \\ \bottomrule
\end{tabular}%
}
\end{table}

\subsection{Procedure}
The study was conducted using a survey on Qualtrics\footnote{\url{https://www.qualtrics.com}}. After collecting informed consent, participants were presented with screenshots of the landing page of four websites in random order and were asked to rate the four items of the VisAWI-s for each website. They were then asked to compare a random pair of the four websites with the comparison item, resulting in 6 comparisons. Participants were also asked to indicate if they knew any of the websites. This procedure was repeated for the second set of four websites, with the order of website categories, websites, and comparisons being randomized. After completion, participants were given the opportunity to leave a comment on the study. \autoref{fig:survey} visualizes the survey procedure.

\subsection{Measures}
To assess the visual aesthetics of the websites, I utilized the VisAWI-s, which consists of 4 items placed below each website screenshot, rated on a 7-point Likert-type scale \parencite[]{moshagenShortVersionVisual2013}. I used the VisAWI-s based on its favorable psychometric properties and its established status as one of the few available survey scales designed to measure website visual aesthetics \parencite{perrig2023visawi}. To minimize participant burden and ensure a unidimensional psychometric structure, the short form of the VisAWI was employed. The use of the long-form VisAWI, with its 18 items and 4 dimensions, would have required a comparison item for each dimension and thus extended the survey beyond feasibility for the present study.
To estimate the smallest subjectively experienced difference, participants were asked to compare each combination of two websites per category in terms of their aesthetics. Specifically, they were asked to indicate whether a website presented on the right was much less, a little less, same, a little more, or much more aesthetically pleasing than a website on the left (see \autoref{fig:anchor}). This comparison item was used to identify participants who perceived a little difference in visual aesthetics between two websites. The wording and answer options were adapted from \textcite{anvariUsingAnchorbasedMethods2021}. 

\section{Results}
\begin{figure*}
    \caption{Mean visual aesthetics rating on the VisAWI-s (N = 249) for the eight websites used as stimuli in the study.}
    \includegraphics[width=\textwidth]{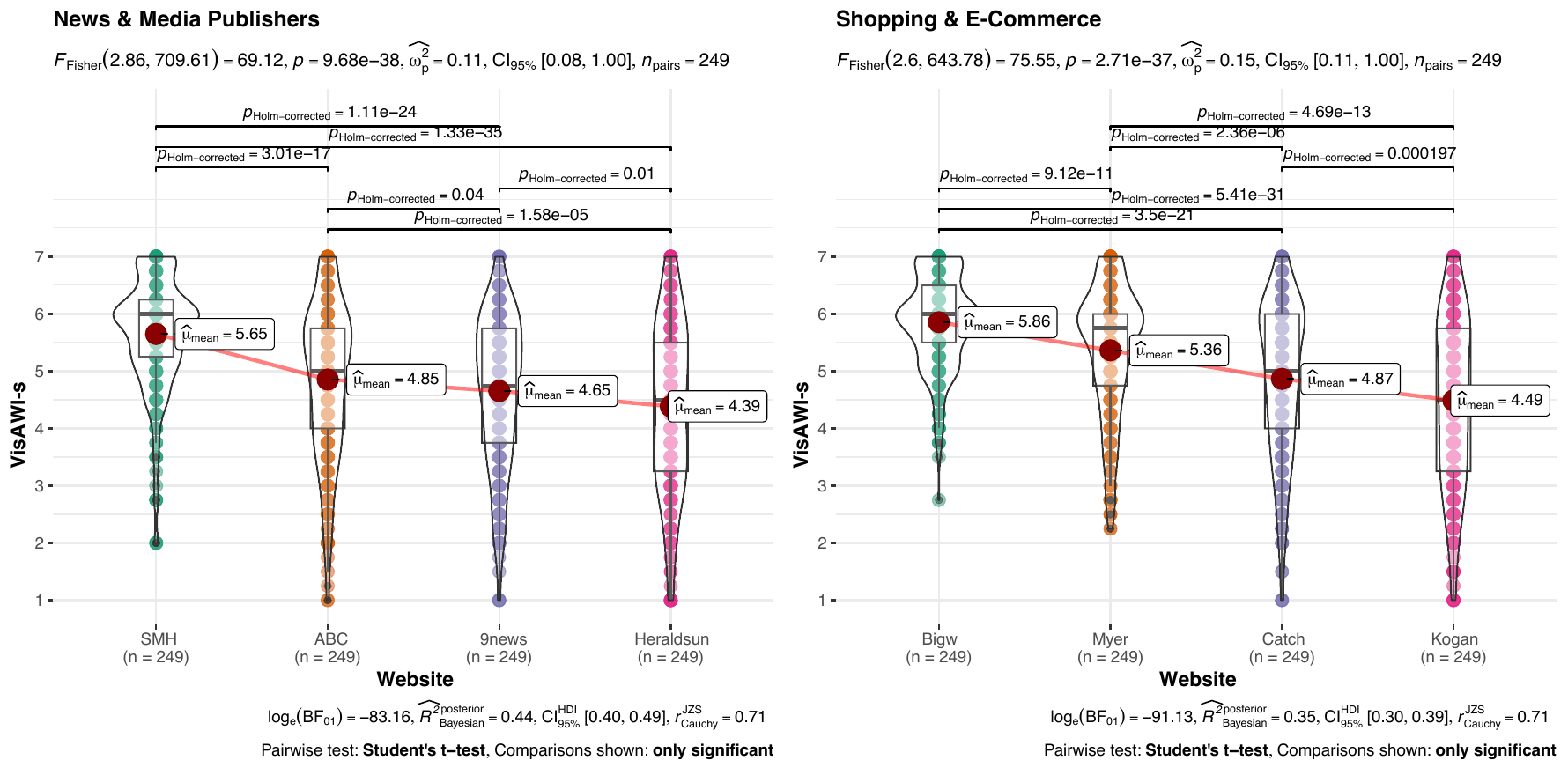}
    \label{fig:overall}
\end{figure*}

Analysis of each website's means in the VisAWI-s revealed several statistically significant differences. See \autoref{fig:overall} for violin plots, including statistical tests, created with the \emph{ggstatsplot} package \parencite{patil2021}. Interestingly, all websites differed statistically significantly using the conventional level of .05 for $\alpha$. The mean rating ranged between 4.39 and 5.65 for news and 4.5 and 5.86 for shopping websites.

For each comparison of two websites, participants who selected "a little more aesthetically pleasing" and "a little less aesthetically pleasing" were included in calculating the smallest subjectively experienced difference (referred to as the "little difference" group in this paper). However, for each pair of websites, one website was rated as statistically significantly higher than the other. Thus, "a little more pleasing" or "a little less pleasing" could go \emph{with} the majority opinion or \emph{against} the majority opinion. I did not have any expert evaluation or ranking of the websites or any other independent, or even an objective (if that even exists), measure to determine which website is actually more aesthetic. Thus, in the first step, I decided not to combine the little difference groups into one but to separately analyze those participants that go \emph{with} or \emph{against} the majority opinion. First, I looked at the standardized mean difference (Cohen's d) and conducted a random effects meta-analysis using \emph{metafor} package \parencite{Viechtbauer2010}. After this, I examined the detailed results for the websites with the smallest overall difference in visual aesthetics ratings for each category. 

\subsection{News \& Media Publisher Websites}
\begin{figure*}
    \caption{Meta-analysis with participants rating a website "a little less" or "a little more" aesthetically pleasing for each comparison of the four news websites. Participants are divided into those whose ratings follow the majority opinion (toward the overall mean) and those whose ratings follow the minority opinion. \emph{Note.} $t_{diff}$ is the t-value associated with the difference between the two websites using ratings from the full sample.}
    \includegraphics[width=\textwidth]{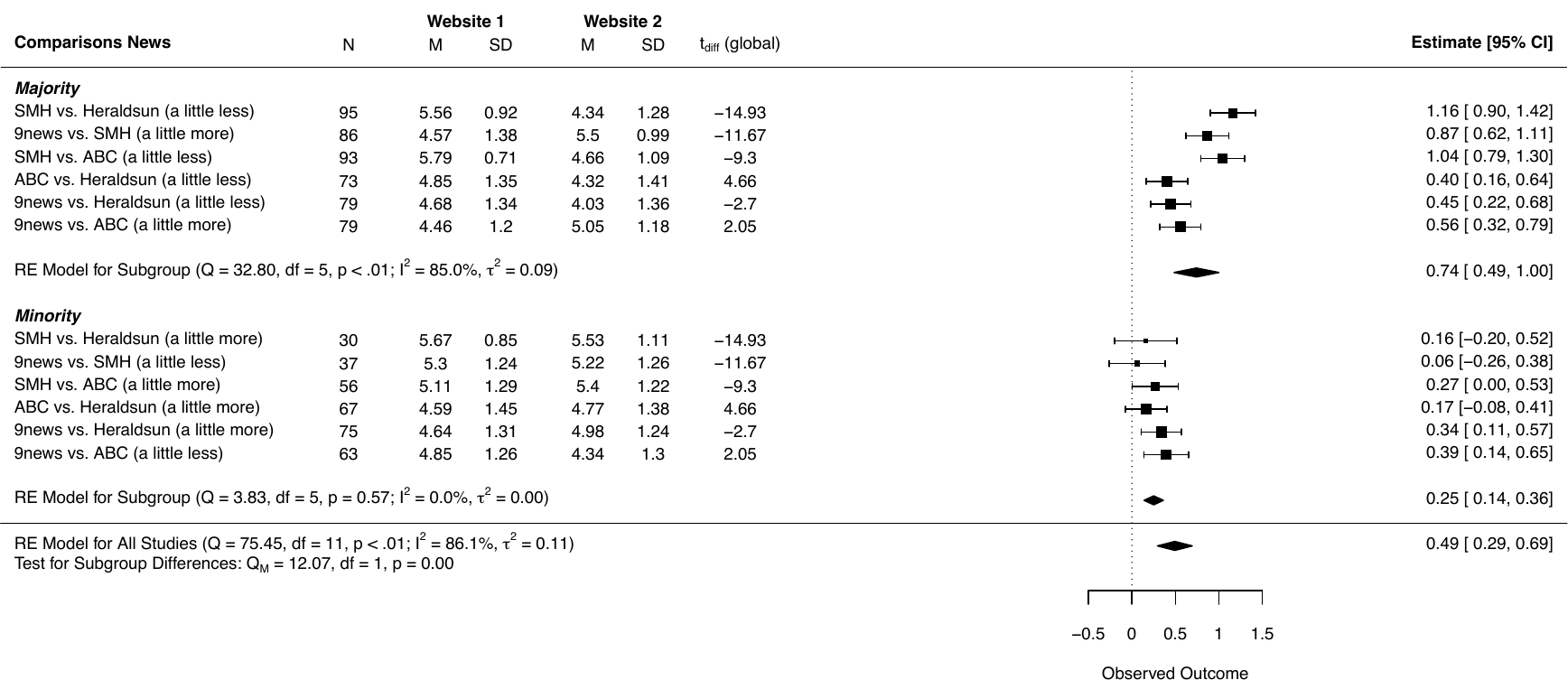}
    \label{fig:news}
\end{figure*}

Results depicted in \autoref{fig:news} suggest that there is substantial heterogeneity over all "little change" groups in the overall model (Q = 75.45, df = 11, $p < .01$; $I^2 = 86.1\%$, $\tau^2 = 0.11$). The majority group showed much larger effect sizes and substantially more heterogeneity.  Given that the overall rating of the websites differed substantially, it can be expected that the effect observed in the little change group was affected by the overall difference between the websites, i.e., meaning that participants experienced a little difference in the direction of the general sentiment reported larger differences if the overall difference between two websites was larger. 

\begin{figure*}
    \caption{Meta-analysis with participants rating a website "a little less" or "a little more" aesthetically pleasing with subgroups for each possible comparison of the news websites. \emph{Note.} $t_{diff}$ is the t-value associated with the difference between the two websites using ratings from the full sample.}
    \includegraphics[width=\textwidth]{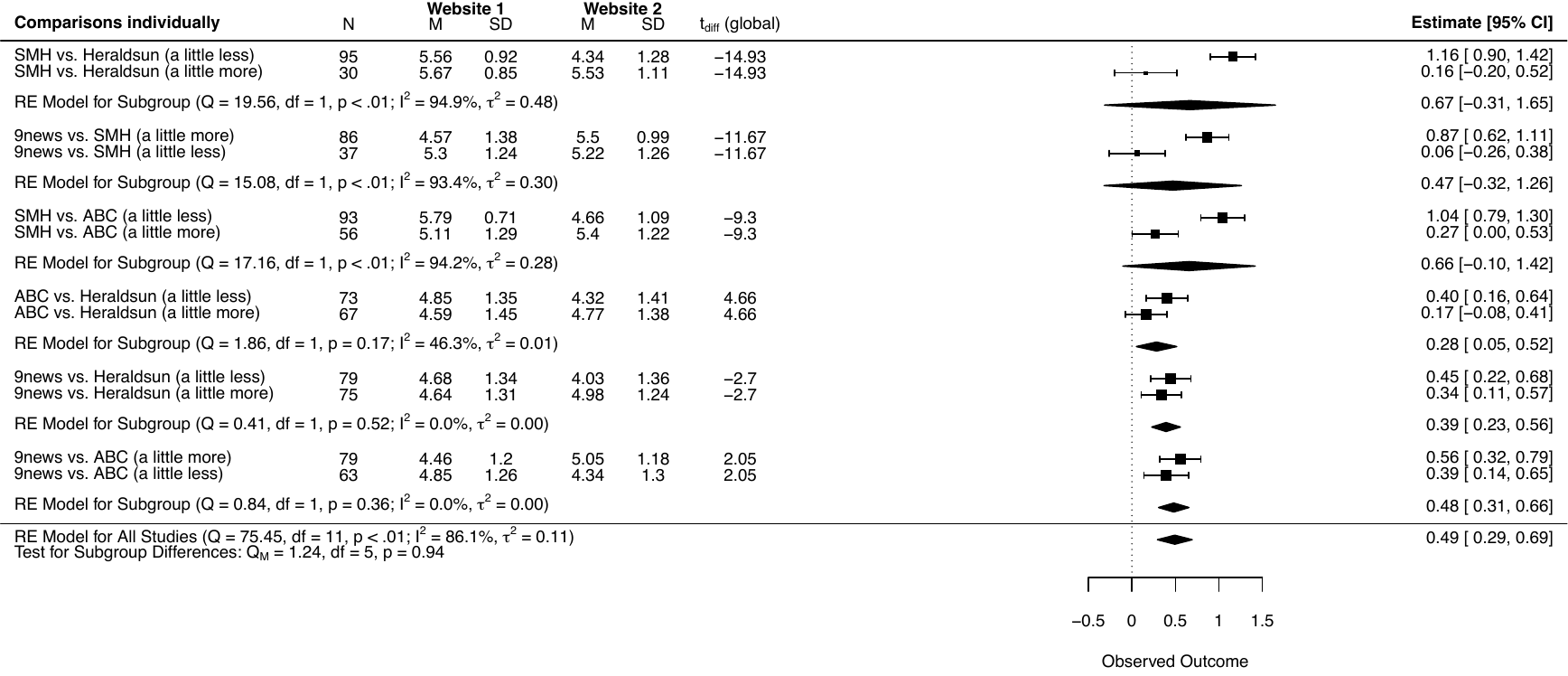}
    \label{fig:newsindiv}
\end{figure*}

Thus, in the next step, I ran the meta-analysis with subgroups for each comparison of websites, including both little difference groups. Results in \autoref{fig:newsindiv} show that for the first three comparisons (SMH vs. Heraldsun, 9news vs. SMH, and SMH vs. ABC), there were quite large differences between the "a little less" and "a little more" groups. The heterogeneity statistics suggest quite large heterogeneity and thus indicate that these estimates should not be combined. However, for websites with similar overall ratings (indicated by smaller absolute $t_{diff}$-values), the two little difference groups could actually be combined as indicated by the non-significant statistics for heterogeneity (i.e., ABC vs. Hearldsun, 9news vs. Hearldsun, and 9news vs. ABC). For websites that are more similar in visual aesthetics, a slight preference for either website is possible. When websites are rated much more differently in visual aesthetics, the ratings of these participants who prefer the less aesthetic website may be less valid due to response errors; or they may have a particular taste. While it could be interesting to explore why certain individuals go against the "mainstream", it seems reasonable that their rating could be less relevant when determining the smallest effect size of interest. 
In \autoref{fig:newsindiv}, it is visible that the three comparisons with the relatively similar websites ABC, Heraldsun and 9news result in effect sizes of similar magnitude. Thus, I decided to run the meta-analysis using only the three comparisons of websites with relatively similar aesthetic ratings.

\begin{figure*}
    \caption{Meta-analysis with participants rating a website "a little less" or "a little more" aesthetically including only the three news websites the most similar overall rating. \emph{Note.} $t_{diff}$ is the t-value associated with the difference between the two websites using ratings from the full sample.}
    \includegraphics[width=\textwidth]{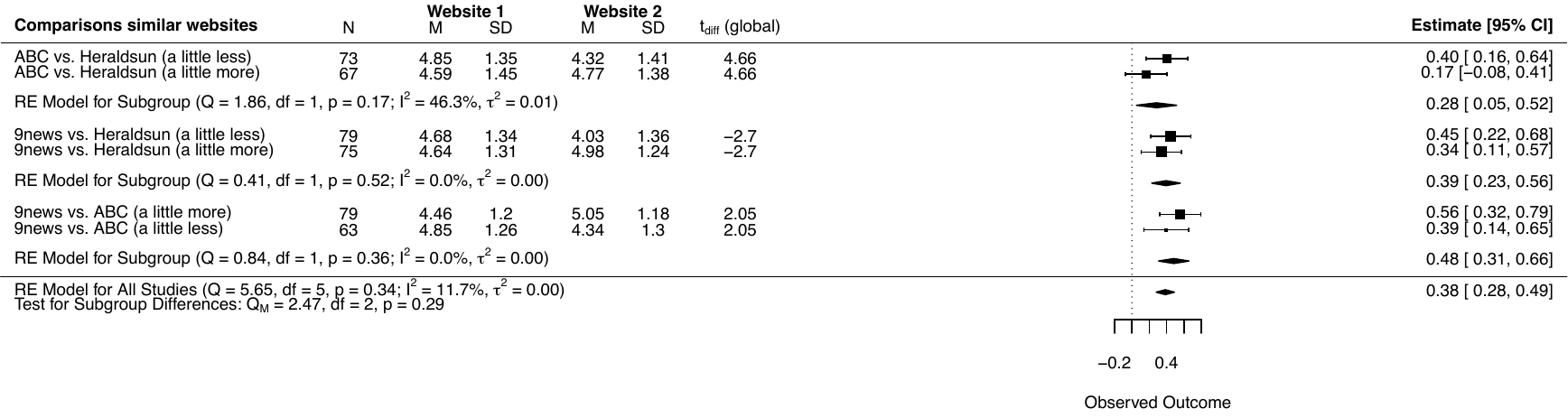}
    \label{fig:newshomo}
\end{figure*}

This analysis's results are depicted in \autoref{fig:newshomo}, which shows no substantial heterogeneity within the subgroups nor over all subgroups. The meta-analytic estimate of the smallest subjectively experienced difference is $d = 0.38$ 95\% CI [0.28, 0.49] for these news websites. 

It would take up too much space in the manuscript to look at the mean difference, effect sizes and POMP for each website comparison. Thus, I decided to just include the detailed analysis for the comparison of the two websites rated most similar in visual aesthetics, 9news and ABC, in this manuscript. \autoref{tab:newsabc} shows the ratings of participants grouped by their answers to the comparison question. The smallest subjectively experienced difference relative to the "same" group is a mean difference of 0.4, $d_{z} = 0.26$, $d_{av} = 0.3$, and a POMP unit change of 6.58\%.
These values could be used to design an informative study of the visual aesthetics of news websites. 

% \usepackage{booktabs}
% \usepackage{graphicx}
\begin{table*}[ht]
\centering
\caption{Means (standard deviations) and mean difference [95\% confidence intervals] in VisAWI-s scores comparing 9news to ABC, with participants subcategorized based on their responses to the comparison item. Cohen's $d_z$ considers the correlations between observations, whereas Cohens's $d_{av}$ does not. Researchers can use Cohen's $d_z$ in power analyses and in equivalence tests. Cohen's $d_{av}$ should be more easily comparable across within- and between-subjects designs.}
\label{tab:newsabc}
\resizebox{\textwidth}{!}{%
\begin{tabular}{@{}lllllS@{}lS@{}lS@{}lS@{}l@{}}
\toprule
\multicolumn{2}{l}{Answer} & N  & 9news: M (SD) & ABC: M (SD) & \multicolumn{2}{c}{Mean Difference}         & \multicolumn{2}{c}{Cohen’s $d_z$}           & \multicolumn{2}{c}{Cohen’s $d_{av}$}           & \multicolumn{2}{c}{POMP}                       \\ \midrule
1 & (much less)            & 24 & 5.48 (0.89)  & 3.74 (1.03) & -1.74 & {[}--2.14;--1.34{]} & -1.84 &{[}--2.5 ;--1.17{]} & -1.81 & {[}--2.45;--1.15{]} & -28.99  & {[}--35.65;--22.34{]} \\
2  & (a little less)           & 63 & 4.85 (1.26)  & 4.34 (1.3)  & -0.51 & {[}--0.83;--0.19{]} & -0.4 & {[}--0.65;--0.14{]} & -0.4 & {[}--0.65;--0.14{]}  & -8.47 & {[}--13.81;--3.12{]}   \\
3 &  (same)           & 33 & 4.89 (1.13)  & 4.78 (1.19) & -0.11 & {[}--0.41;  0.18{]}  & -0.14 & {[}--0.48; 0.21{]} & -0.1 & {[}--0.35; 0.15{]}   & -1.89 & {[}--6.86; 3.08{]}     \\
4  & (a little more)            & 79 & 4.46 (1.2)   & 5.05 (1.18) & 0.59 & {[}  0.35; 0.82{]}    & 0.56 & {[} 0.32; 0.8{]}    & 0.49 & {[} 0.28; 0.7{]}     & 9.81 & {[} 5.9; 13.72{]}       \\
5 & (much more)            & 49 & 4.19 (1.61)  & 5.88 (1.09) & 1.68 & {[} 1.23; 2.13{]}    & 1.08 & {[} 0.72; 1.42{]}   & 1.23 & {[} 0.82; 1.63{]}    & 28.06 & {[} 20.57; 35.55{]}    \\ \bottomrule
\multicolumn{3}{@{}l}{\small \textit{Note.} Total $N$ = 248.}\\
\end{tabular}
}
\end{table*}

\subsection{Shopping \& E-commerce Websites}
Similar to the news category, comparisons of shopping websites showed substantial heterogeneity overall ($Q = 41.20$, $df = 11$, $p < .01$; $I^2$ = 75.0\%, $\tau^2 = 0.06$). Thus, I looked again at the subgroups for participants in the little change groups that follow the majority or the minority opinion in the comparison of two websites. Figure 5 shows that there is a lot of heterogeneity in the majority group, but less so in the minority group. 

\begin{figure*}
    \caption{Meta-analysis with participants rating a website "a little less" or "a little more" aesthetically pleasing for each comparison of the four shopping websites. Participants are divided into those whose ratings follow the majority opinion (toward the overall mean) and those whose ratings follow the minority opinion. $t_{diff}$ is the t-value associated with the difference between the two websites using ratings from the full sample.}
    \includegraphics[width=\textwidth]{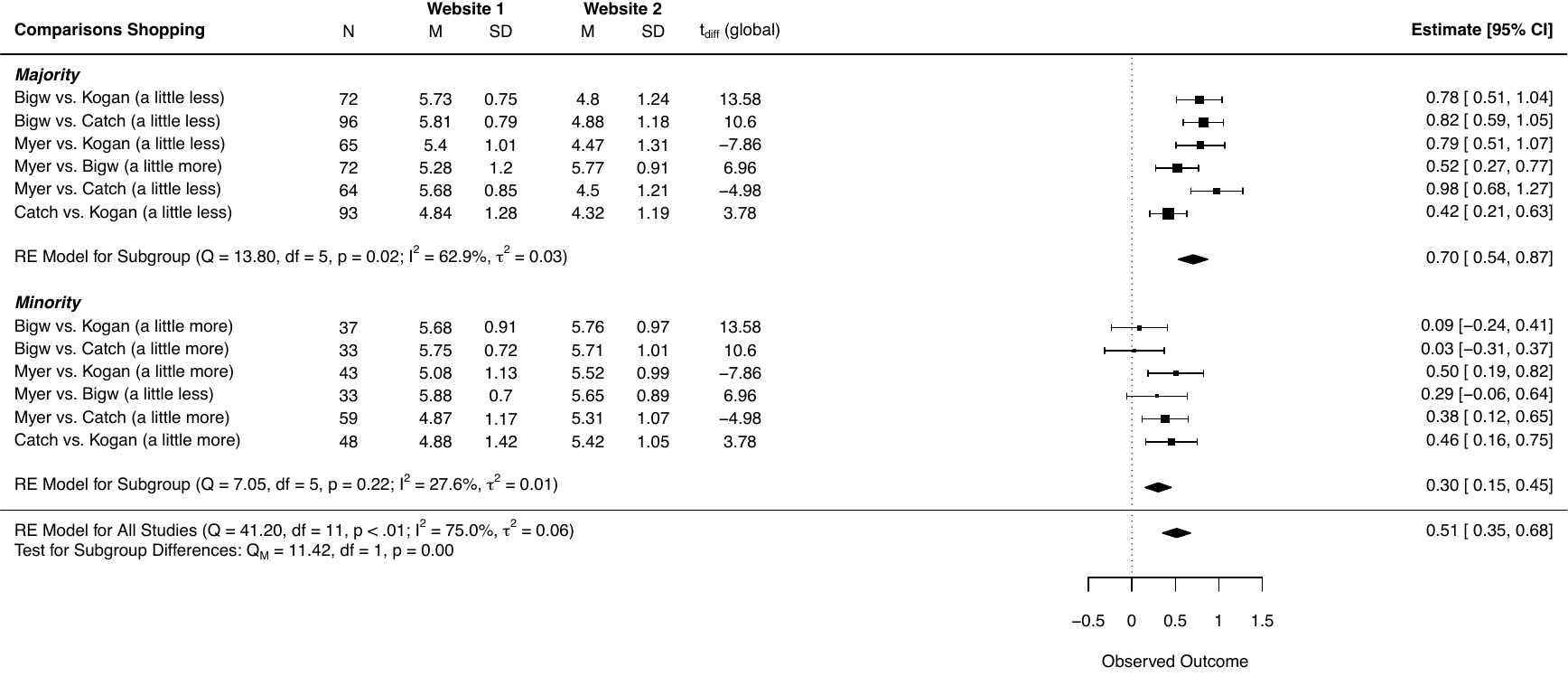}
    \label{fig:shop}
\end{figure*}

Figure 6 depicts the analysis for each comparison between two websites individually. It is again visible that websites with large differences in the overall ratings of participants have a lot of heterogeneity in their little difference groups, indicating that participants who go against the majority opinion in the comparison differ substantially from those who go with the majority. Interestingly, the heterogeneity was also quite substantial in the comparison between Myer and Catch, which showed smaller differences in overall ratings than Myer and Kogan. 

\begin{figure*}
    \caption{Meta-analysis with participants rating a website "a little less" or "a little more" aesthetically pleasing with subgroups for each possible comparison of the shopping websites. $t_{diff}$ is the t-value associated with the difference between the two websites using ratings from the full sample.}
    \includegraphics[width=\textwidth]{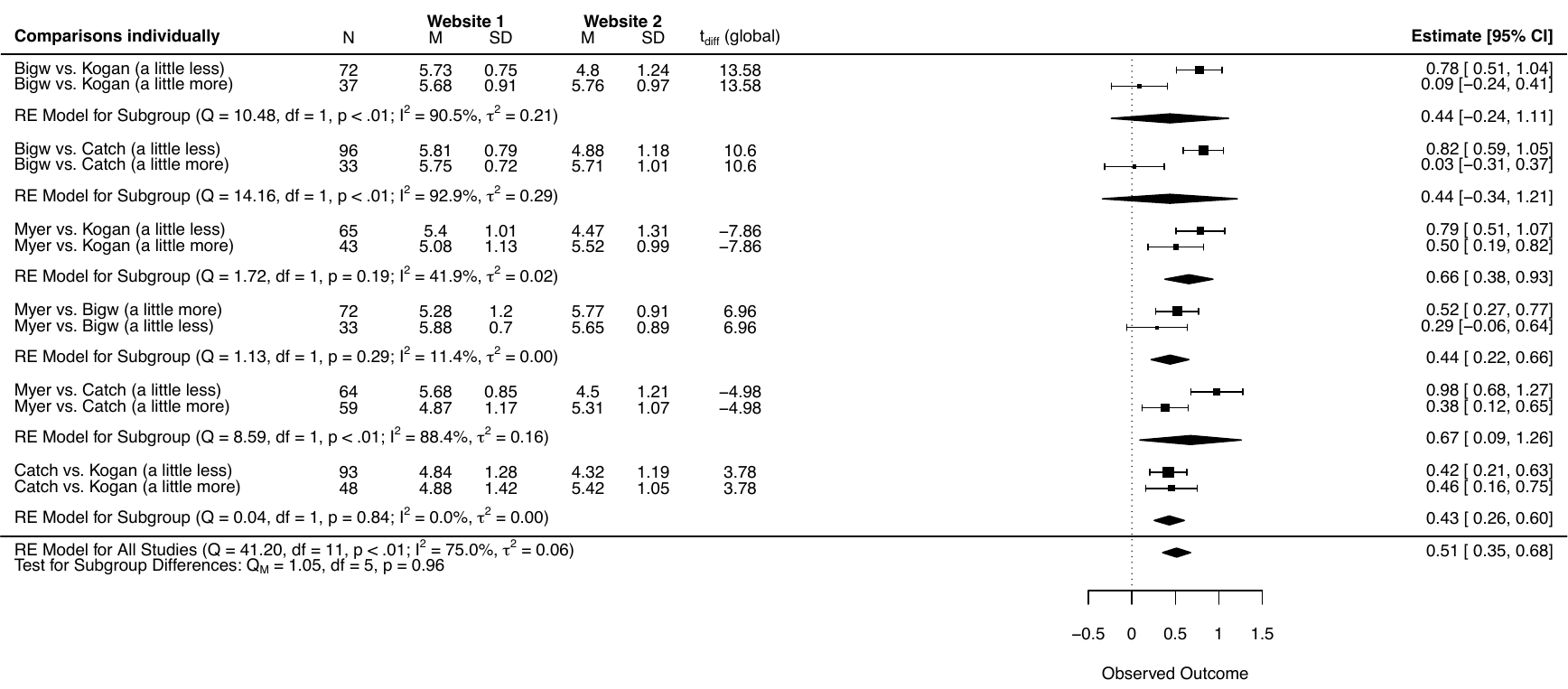}
    \label{fig:shopindiv}
\end{figure*}

As with the news websites, I decided to run the meta-analysis using only the comparison with low heterogeneity, which might indicate that a participant's decision for a website could go either way. These websites are more similar in their overall visual aesthetics rating (with the exception above). Figure 7 depicts no substantial heterogeneity within the subgroups and over all comparisons. The meta-analytic estimate of the smallest subjectively experienced difference is d = 0.50 95\% CI [0.38, 0.62] for these shopping websites. 

\begin{figure*}
    \caption{Meta-analysis with participants rating a website "a little less" or "a little more" aesthetically pleasing including only the three shopping websites the most similar overall rating. $t_{diff}$ is the t-value associated with the difference between the two websites using ratings from the full sample.}
    \includegraphics[width=\textwidth]{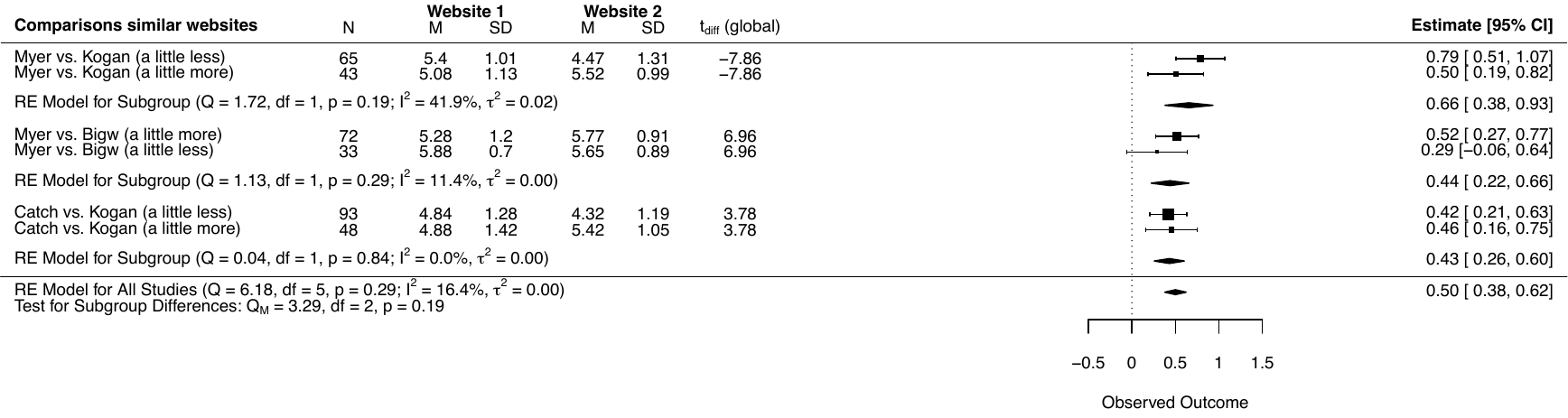}
    \label{fig:shophomo}
\end{figure*}

I include a detailed analysis to compare the two websites rated most similar in visual aesthetics, Catch and Kogan. \autoref{tab:catchkogan} shows participants' ratings grouped by their answer on the comparison question. The smallest subjectively experienced difference relative to the "same" group is a mean difference of 0.52, $d_z = 0.44$, $d_av = 0.42$, and POMP unit change of 8.56\%. Again, these values could be used to design an informative study of the visual aesthetics of shopping websites.

% \usepackage{booktabs}
% \usepackage{graphicx}
\begin{table*}[ht]
\centering
\caption{Means (standard deviations) and mean difference [95\% confidence intervals] in VisAWI-s scores comparing Catch to Kogan, with participants subcategorized based on their responses to the comparison item.}
\label{tab:catchkogan}
\resizebox{\textwidth}{!}{%
\begin{tabular}{@{}lllllS@{}lS@{}lS@{}lS@{}l@{}}
\toprule
\multicolumn{2}{l}{Global Answer} & N  & Catch: M (SD) & Kogan: M (SD) & \multicolumn{2}{c}{Mean Difference}         & \multicolumn{2}{c}{Cohen's $d_z$}           & \multicolumn{2}{c}{Cohen's $d_{av}$}           & \multicolumn{2}{c}{POMP}                       \\ \midrule
1 & (much less)            & 61 & 5.05 (1.36)  & 3.33 (1.32) & -1.72 & {[}--2.08;--1.36{]} & -1.23 &{[}--1.56 ;--0.89{]} & -1.28 & {[}--1.63;--0.93{]} & -28.69  & {[}--34.66;--22.72{]} \\
2  & (a little less)           & 93 & 4.84 (1.28)  & 4.32 (1.19)  & -0.52 & {[}--0.78;--0.27{]} & -0.42 & {[}--0.63;--0.21{]} & -0.42 & {[}--0.63;--0.21{]}  & -8.69 & {[}--12.94;--4.45{]}   \\
3 &  (same)           & 22 & 4.42 (1.56)  & 4.44 (1.67) & 0.02 & {[}--0.49;  0.54{]}  & 0.02 & {[}--0.04; 0.44{]} & 0.01 & {[}--0.29; 0.31{]}   & 0.38 & {[}--8.21; 8.97{]}     \\
4  & (a little more)            & 48 & 4.88 (1.42)   & 5.42 (1.05) & 0.54 & {[}  0.2; 0.87{]}    & 0.46 & {[} 0.16; 0.76{]}    & 0.43 & {[} 0.15; 0.7{]}     & 8.94 & {[} 3.33; 14.56{]}       \\
5 & (much more)            & 25 & 4.86 (1.43)  & 6.19 (0.58) & 1.33 & {[} 0.81; 1.85{]}    & 1.06 & {[} 0.56; 1.54{]}   & 1.22 & {[} 0.65; 1.78{]}    & 22.17 & {[} 13.53; 30.81{]}    \\ \bottomrule
\multicolumn{3}{@{}l}{\small \textit{Note.} Total $N$ = 248.}\\
\end{tabular}
}
\end{table*}

\section{Discussion}
Determining whether a difference in visual aesthetics between websites is theoretically and practically meaningful is crucial for website designers and researchers. This study proposes an approach to this problem using the anchor method presented in \textcite{anvariUsingAnchorbasedMethods2021}. Unlike the example in \textcite{anvariUsingAnchorbasedMethods2021}, where participants reported changes in their affect over time, participants in this study rated four News \& Media publishers websites and four Shopping \& E-commerce websites and compared all websites within a category.

I calculated estimates for the relative difference, compared to those participants who report "the same" visual aesthetics, that led participants to reported "a little difference" for news (M = 0.4) and "a little difference" for shopping websites (M = 0.52). These estimates, together with their standard deviations, can be used as the smallest effect size of interest in a-priori power analyses for studies that use the VisAWI-s in the population my sample was drawn from (i.e., US participants from Prolific). Alternatively, the smallest subjectively experience difference in POMP units, which can be compared across contexts and measures, were: 6.58\% for news and 8.56\% for shopping websites. This means, using the original VisAWI-s scale with average scoring (range from 1 to 7), a difference above 7 × 6.58\%  = 0.46 for news websites or 7 × 8.56\% = 0.60 shopping websites is a subjectively noticeable difference.  

It is important to note, though, that the associated difference in VisAWI-s ratings for websites that participants consider "a little more" or "a little less" aesthetically is dependent on the stimuli used. When the difference between the overall VisAWI-s rating was large, participants who only reported "a little more" or "a little less" tended to report larger differences in their rating. Thus, it seems that the most meaningful benchmark of a smallest subjectively experienced difference should be determined with two websites that are relatively similar in their overall visual aesthetics. However, taken to the extreme, if participants would rate exactly the same website twice and report "a little difference," then the associated effect size would be meaningless \emph{or} it may serve as a lower bound of effect sizes. In the present study, all websites differed; thus, the resulting smallest subjectively experienced difference in VisAWI-s ratings is meaningful, but it could be overestimated. This is the reason why it is important to use this method to determine the smallest effect size of interest for the measure you use in the populations you study.

However, I believe that the following interpretation is permitted: The reported effect size (or mean difference and POMP, respectively) can be used as a threshold, meaning that an observed difference \emph{equal or above} this effect size is subjectively experienced or noticeable by users. However, the interpretation that a difference \emph{below} this effect size is not relevant or not noticeable by users is not permitted. 
This means that, while I aimed to determine a benchmark to decide whether a difference in VisAWI-s ratings between two websites is meaningful, I may not be able to provide a definitive answer with this study alone. However, I can provide a (potentially overestimated) threshold for meaningful effect size differences of the VisAWI-s, which is minimally 0.46 on a 7-point Likert scale or as low as 0.4 when using a US-based sample from Prolific with news websites from Australia. Returning to the differences in overall mean ratings for the eight websites depicted in Figure 2, using the POMP-based threshold of 0.46, the differences between SMH and the other news websites are definitely noticeable (mean difference from 0.8 to 1.26) and between ABC and Heraldsun (mean difference 0.46). The meaning of the other observed significant differences between the news websites may be less important. 
For the shopping websites, all mean differences exceed the threshold, except for the difference between Catch and Kogan (mean difference 0.38). The interpretation that this (statistically significant) difference is not practically and/or theoretically meaningful may, however, not be permitted because, with the present study, the smallest subjectively experienced difference might have been overestimated. 

\section{Conclusion}
This paper reports an exploratory study of meaningful effect sizes in ratings of visual aesthetics of websites using the VisAWI-s. Results show that the selection of the stimuli affect potential estimates of a smallest subjectively experienced difference. Using the comparison of the two most similar websites in users' overall visual aesthetics ratings (9news and Heraldsun), I determined an estimate of a population-specific mean difference of 0.4, or in POMP units 6.58\%, which translates to a mean difference of 0.46 with the 7-point Likert-type scale of the VisAWI-s. These values may be used to inform research on visual aesthetics of websites. If researchers observe differences exceeding these estimates, then it is likely that these differences are meaningful. However, if an observed difference undercuts this estimate, it may not be permitted to classify such a difference as not meaningful. Instead, future research needs to carefully calibrate such a threshold to determine whether two websites can be considered equal in their perceived visual aesthetics.   

\section*{Acknowledgements}
Special thanks to my wonderful team: Lena, Sebastian, Nicolas, Melanie, Zgjim, Ariane, Nick, Antony, Léane, and--of course--Klaus. 

\printbibliography

\appendix

% Screenshots of website
% Anchor item used
% VisAWI-s
% Link to OSF repository

\section{Instrument and Data Availability}

\begin{figure}
    \centering
    \caption{Example of the anchor item used in the study.}
    \includegraphics[width=0.45\textwidth]{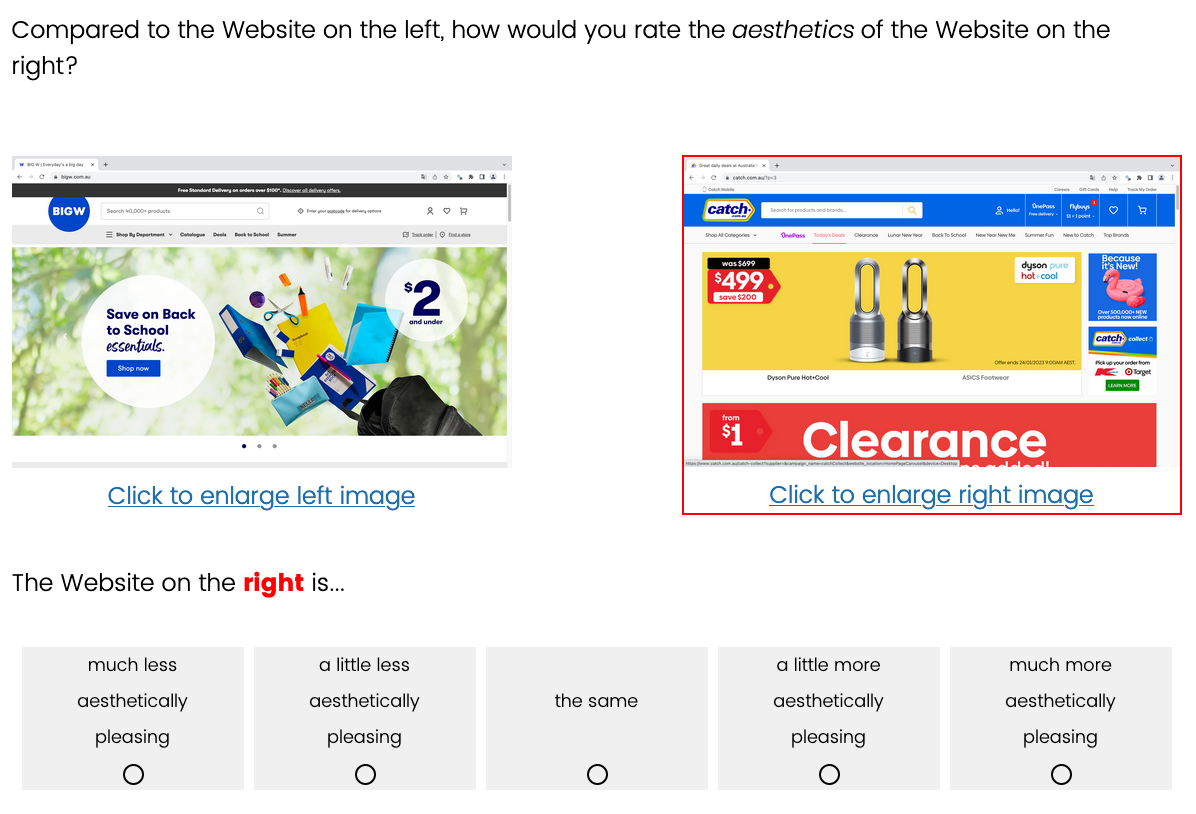}
    \label{fig:anchor}
\end{figure}

\begin{figure*}
    \centering
    \caption{Screenshots of the websites on 2023-01-19 used as stimuli in the study.}
    \includegraphics[width=0.49\textwidth]{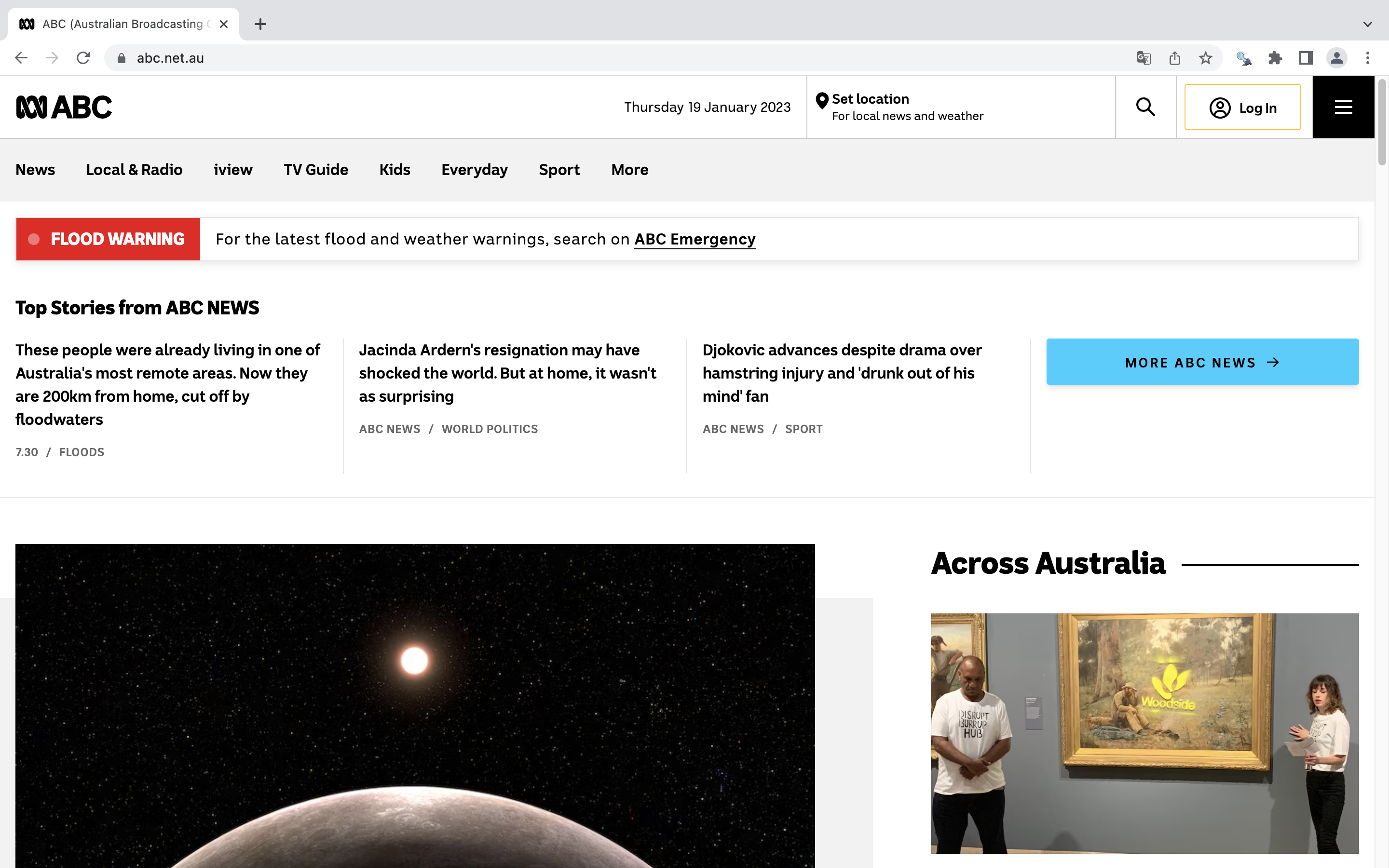}
    \includegraphics[width=0.49\textwidth]{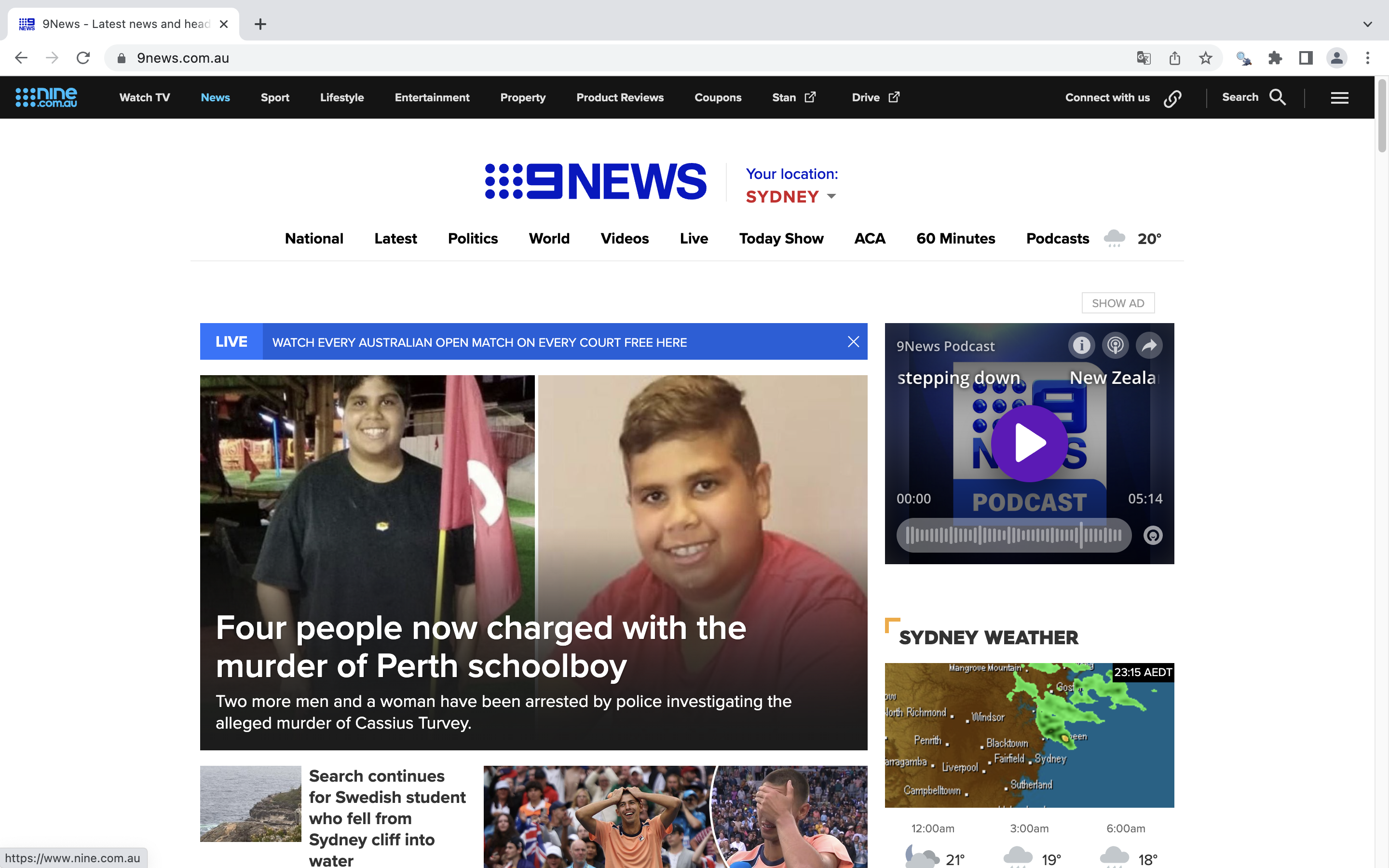}
    \includegraphics[width=0.49\textwidth]{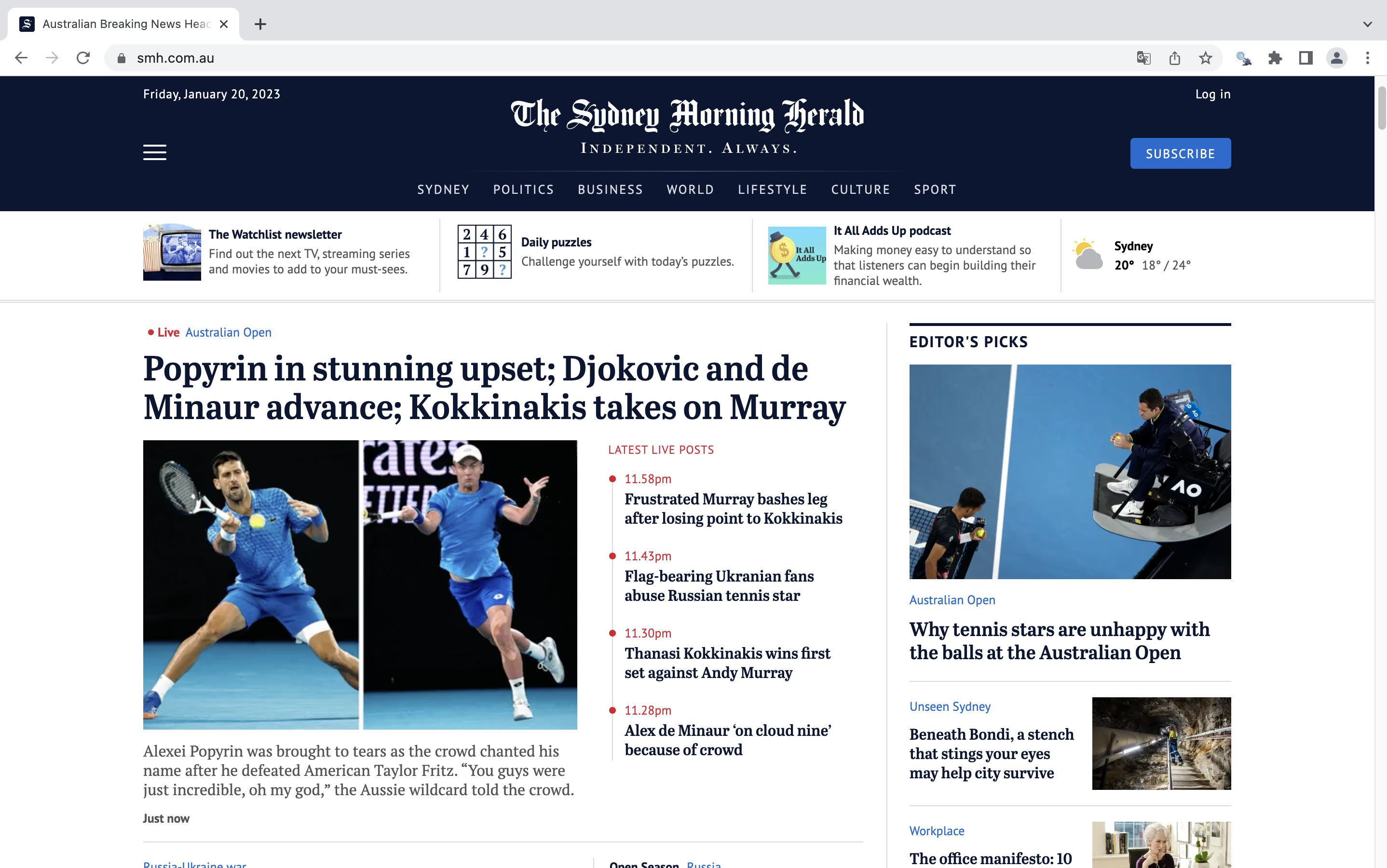}
    \includegraphics[width=0.49\textwidth]{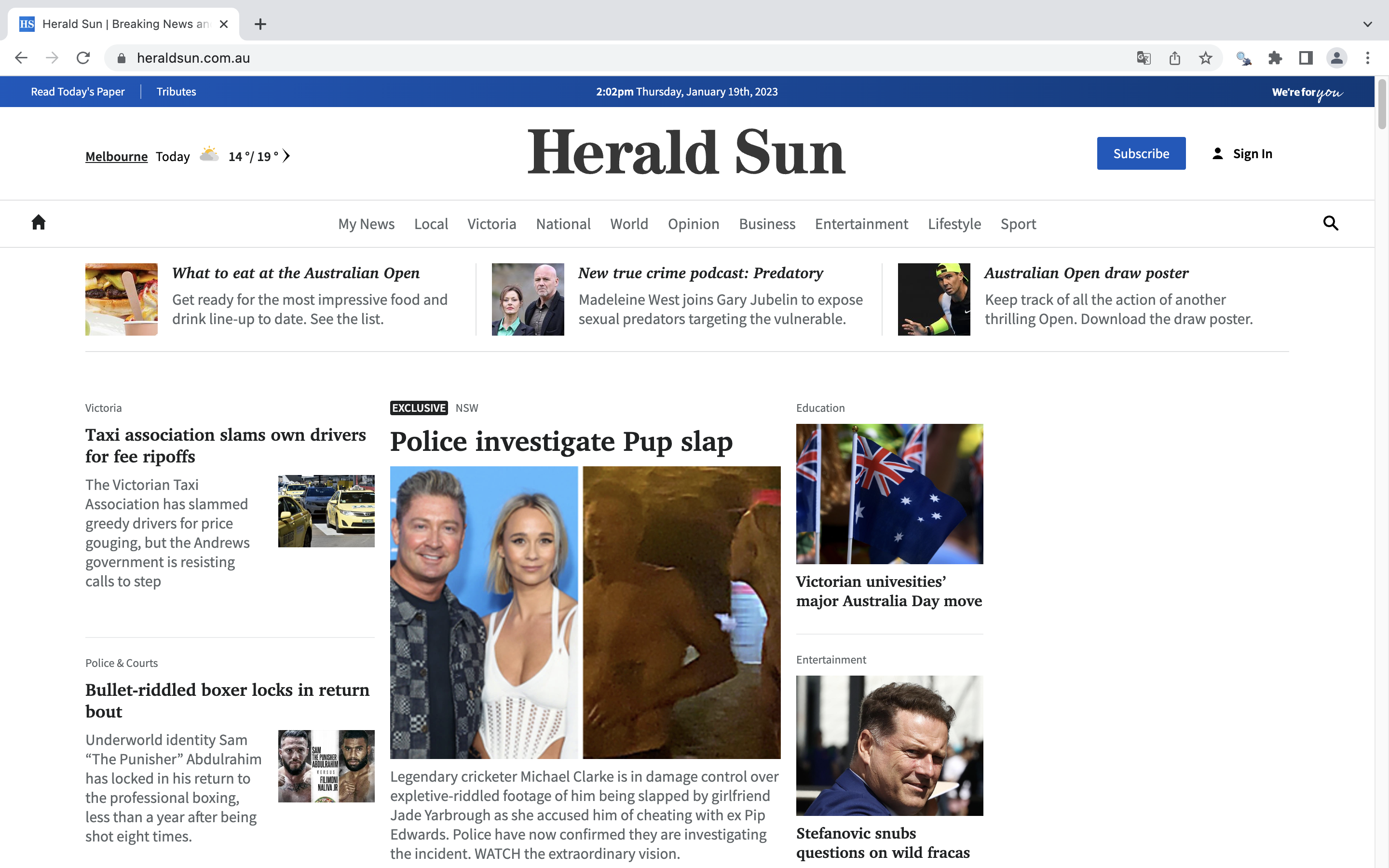}
    \includegraphics[width=0.49\textwidth]{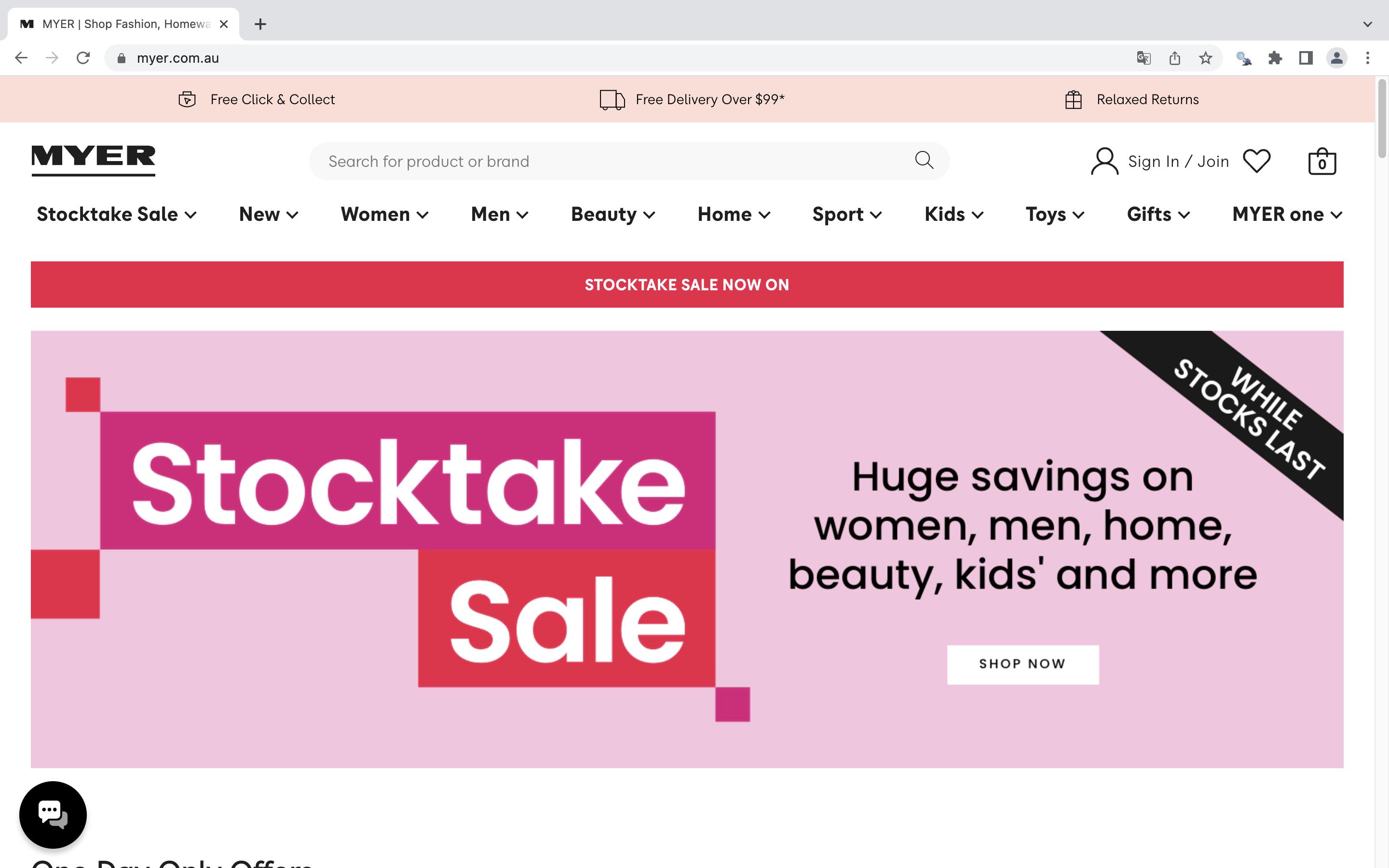}
    \includegraphics[width=0.49\textwidth]{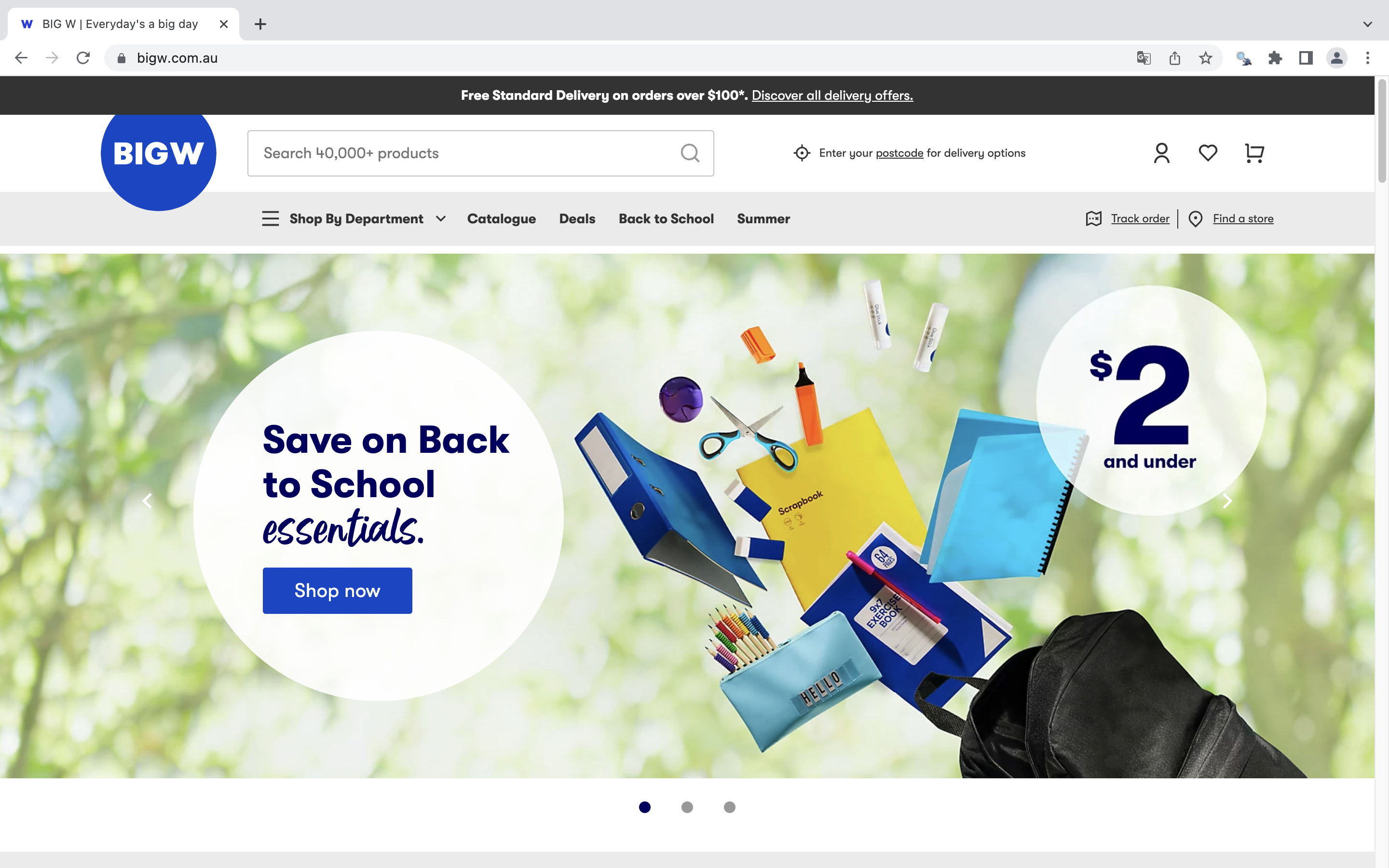}
    \includegraphics[width=0.49\textwidth]{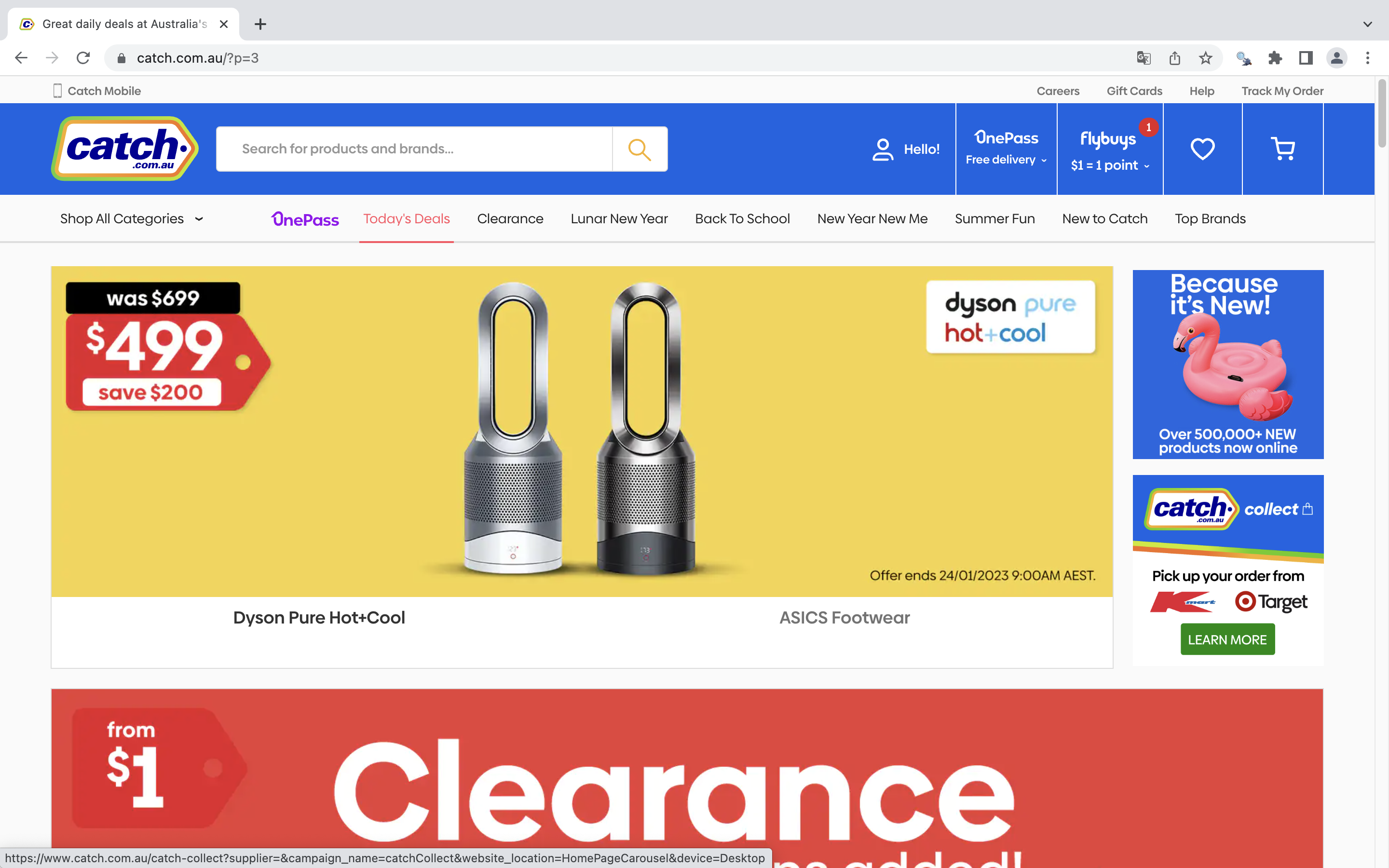}
    \includegraphics[width=0.49\textwidth]{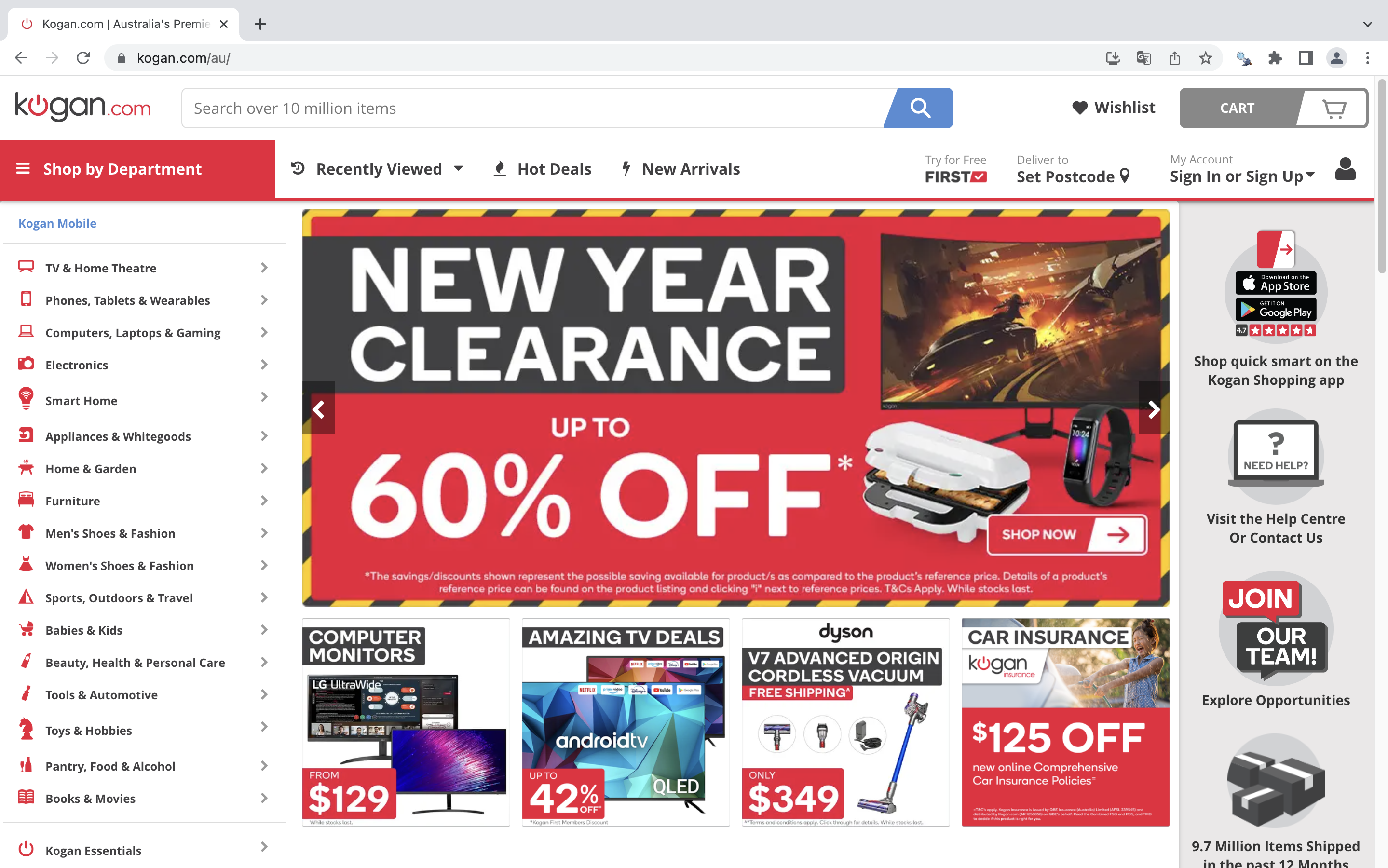}
    \label{fig:websites}
\end{figure*}

All screenshots used in this study are presented in \autoref{fig:websites}. The comparison item, as implemented on the survey platform, is depicted in \autoref{fig:anchor}. \autoref{tab:visawis} shows all four items of the VisAWI-s used in this study. 

Data and analysis code are available on OSF (\url{https://osf.io/ceydz/?view_only=b2db0ee6feee4f719b36ad66fc6d7bcb}).

\begin{table}[h]
\centering
\caption{Items of the VisAWI-s were presented with a 7-point Likert-type agreement scale ranging from "strongly disagree" to "strongly agree." }
\label{tab:visawis}
\begin{tabular}{@{}ll@{}}
\toprule
No.  & Item    \\ \midrule
1 & The layout appears professionally designed.   \\
2 & The color composition is attractive.   \\
3 & The layout is pleasantly varied.   \\
4 & Everything goes together on this site.   \\
 \bottomrule
\end{tabular}%
\end{table}

\begin{figure*}
    \centering
    \caption{Flowchart of the survey procedure.}
    \includegraphics[width=0.98\textwidth]{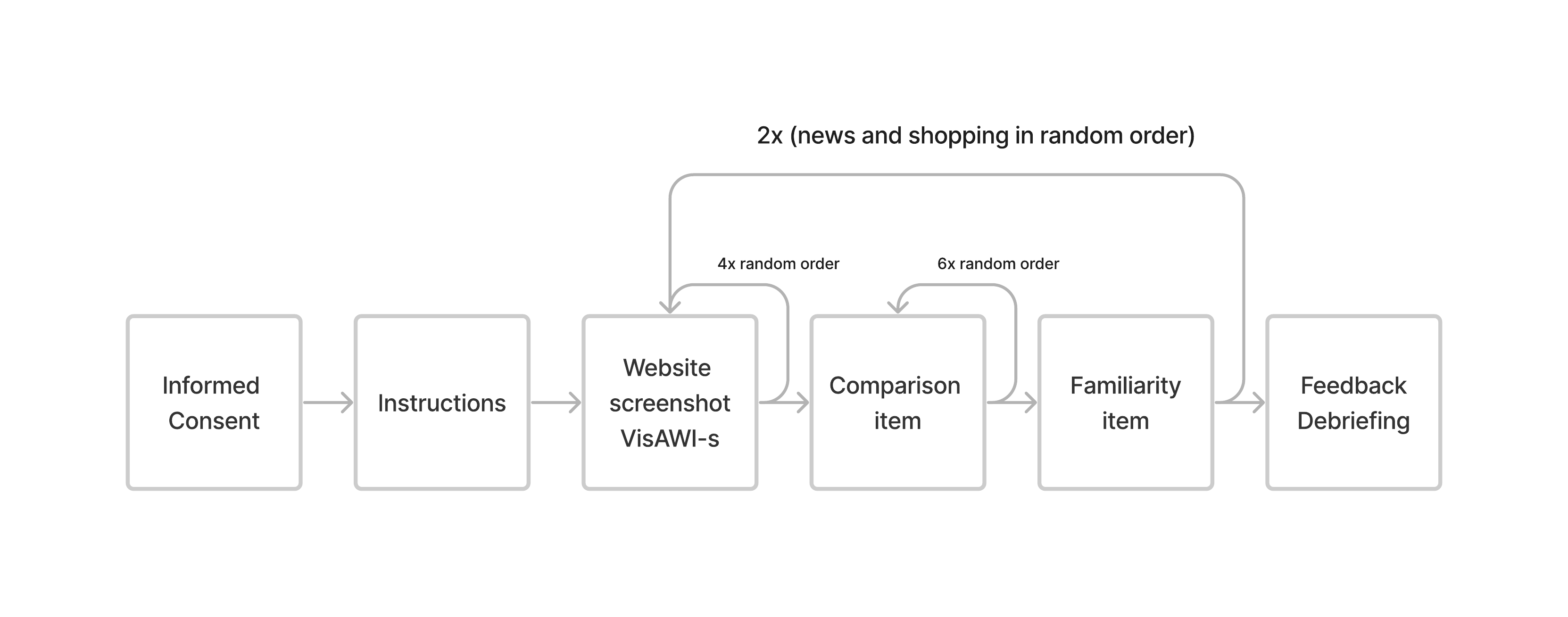}
    \label{fig:survey}
\end{figure*}

\end{document}